\documentclass[11pt]{elsarticle}

\usepackage{graphicx,amsmath,amssymb,latexsym,epsfig,epstopdf}
\usepackage{bbm}
\usepackage{subfigure}

\usepackage{a4wide}
\newtheorem{lemma}{Lemma}

\newtheorem{theorem}{Theorem}
\newtheorem{prop}{Proposition}

\def\I{\mathbbm{1}}

\def\bp{\noindent{\it Proof.}\ }
\def\ep{\hfill $\Box$}
\def\sumi{\sum_{i=1}^N}
\def\sumk{\sum_{k=1}^K}

\def\N{{\mathbb N}}
\def\et{{\quad \mbox{and}\quad }}

\begin{document}

\begin{frontmatter}

\title{On the stability of  flow-aware CSMA}

\author[tpt]{T.~Bonald}
\author[inria]{M.~Feuillet}
\address[tpt]{Telecom ParisTech, Paris, France}
\address[inria]{INRIA, Rocquencourt, France}

\begin{abstract}
We consider a wireless network where each {\it flow} (instead of each link)  runs its own CSMA (Carrier Sense Multiple Access)
algorithm. Specifically, each  flow attempts to access the radio  channel
after some random time and transmits a packet if the channel is sensed idle. We
prove that, unlike the standard CSMA algorithm, this simple distributed access scheme is optimal
in the sense that the network is stable   for all traffic intensities  in the
capacity region of the network.
\end{abstract}


\end{frontmatter}

\section{Introduction}
  
\begin{figure}[h]
\begin{center}
\subfigure[Downstream vs.~{upstream}]{\includegraphics[width=4cm]{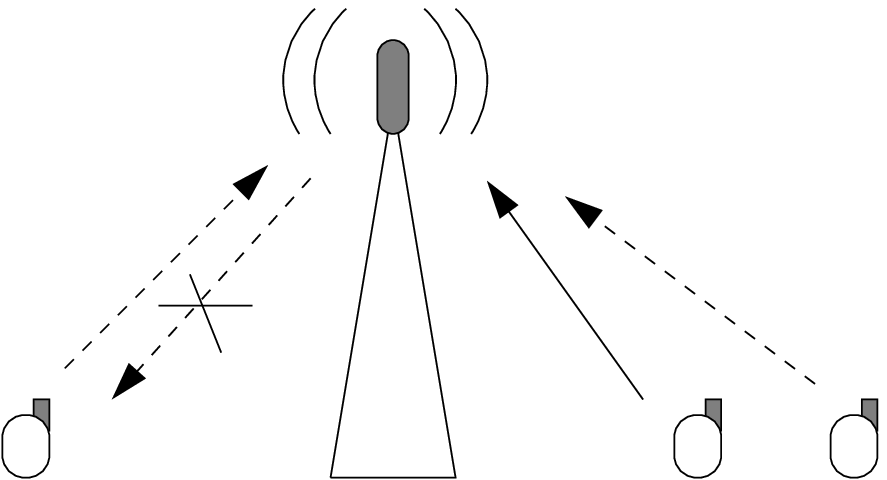}}
\hspace{2cm}
\subfigure[Interference]{\includegraphics[width=7cm]{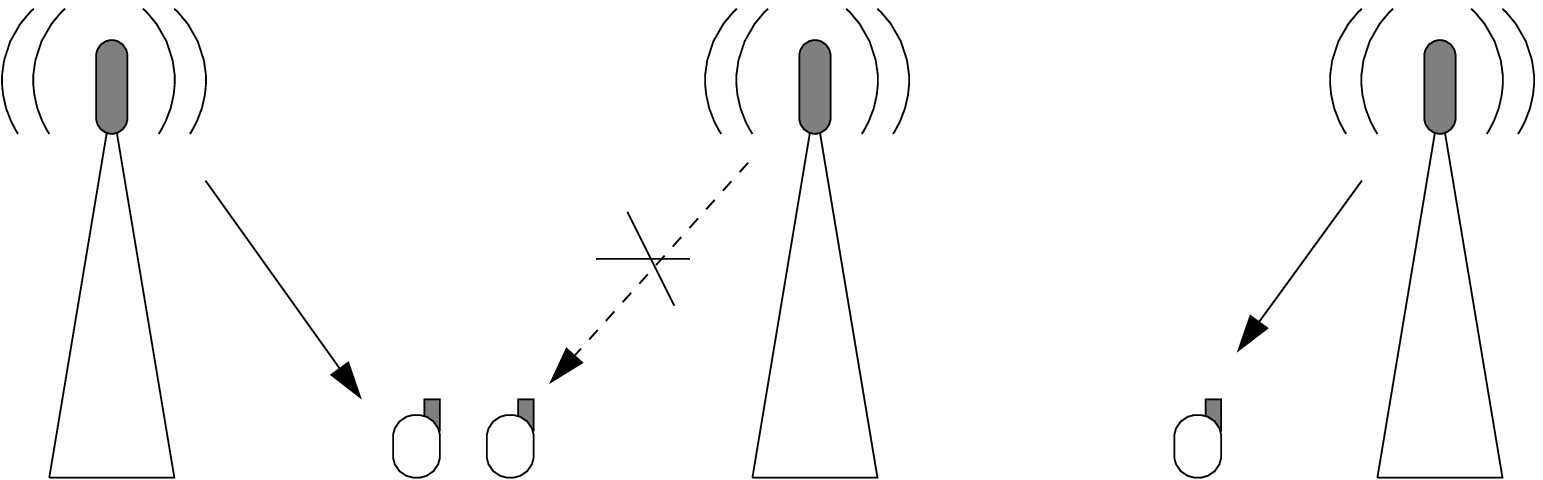}}
\caption{\label{tnet}Unfairness of standard CSMA.}
\end{center}
\end{figure}

 The CSMA (Carrier Sense Multiple Access) algorithm is one of the most common medium  access schemes in today's networks, both wired (e.g.~IEEE 802.3) and wireless (e.g.~IEEE 802.11). However, this algorithm is known to be inherently unfair, as illustrated by the two scenarios of Fig.~\ref{tnet}.  
The first scenario relates to the downstream vs. upstream bandwidth sharing for a single access point.
In the presence of $n$ active mobiles on the upstream, the access point competes with $n$ nodes for accessing the channel, resulting in a downstream to upstream bandwidth ratio of $1/n$, independently on the number of active flows on the downstream. The second scenario illustrates the impact of interference on bandwidth sharing. 
 The center access point cannot transmit if one of the edge access points is active and thus gets much less transmission opportunities. 
 Moreover, the resulting bandwidth sharing is inefficient since the edge access points can access the channel alternately, preventing the center access point from sending its traffic.  Thus the  CSMA algorithm is not able to fully utilize network  capacity, a statement that will be made  more precise later in the paper.

 We propose a slight modification of the standard CSMA algorithm that consists in running the algorithm for each {\it flow} instead of each transmitter. In this paper, we refer to a flow as any file transfer from a source to a destination; it can typically be identified through the usual 5-uple: IP source and destination addresses, source and destination ports, protocol. For a single access point, each {\it flow} (either downstream or upstream) runs the CSMA algorithm and thus gets the same bandwidth share. The whole system can then be viewed as a unique, evenly shared wireless link. The focus of the present paper is rather on the second scenario where some links suffer from high interference. Specifically, we show that the flow-aware CSMA algorithm is optimal in the sense that it stabilizes the network whenever possible. 
 In the example of Fig.~1, the center access point is likely to access the channel when it has a high number of active flows; at the end of the corresponding activity period, the edge access points can access the channel and will likely be {\it simultaneously} active, which is a necessary condition for fully utilizing network capacity.
 
 The main result of the paper is to demonstrate that the flow-aware CSMA algorithm is optimal for {\it any}  network topology.  We consider a general model consisting of an arbitrary number of wireless links whose mutual interference is represented by some conflict graph. Flows of random size arrive at random at each link. In order to study the flow-level dynamics, we calculate the throughput of each flow granted by the CSMA algorithm under the usual time-scale separation assumption. We then prove that, provided there exists some schedule of the links that stabilizes the network, the flow-aware CSMA algorithm will do so, in a purely distributed and asynchronous way. 
   
The rest of the paper is organized as follows. Related work is presented in the next section. We then present the model and analyse its stability under standard and flow-aware CSMA, respectively. The impact of network load on the mean throughput of each flow under flow-aware CSMA is considered in Section \ref{sec:perf}. Section \ref{sec:conclusion} concludes the paper.

\section{Related work}
The problem of optimal bandwidth sharing in wireless networks has first been tackled by Tassiulas and Ephremides, who showed in \cite{tassiulas-92} that the so-called {\it maximal weight} scheduling policy, which activates a set of links that  maximizes the total backlog of active links, stabilizes any network whenever possible. A number of distributed implementations of this policy have then been proposed, all relying on some message passing protocol between nodes, see e.g.~\cite{modiano-06,raja-09}. Simple heuristics based on greedy algorithms that require limited or no message passing have also been studied, most selecting schedules of {\it maximal  size} (in terms of number of links) instead of maximal weight  and, as such, being suboptimal \cite{dimakis-06,gupta-09,joo-09,leconte-09,proutiere-08,wu-07}.

A new approach to optimal scheduling has recently been proposed by Jiang and Walrand, who introduced in \cite{walrand-08} a distributed CSMA algorithm where at each link, the attempt rate  is adapted to the arrival rate and service rate so as to meet the demand.
The result is based on a time-scale separation assumption whereby the activity states of the links, which  depend on   the CSMA algorithm, evolve much faster than the attempt rates of the links. In practice, the algorithm used for adapting the attempt rates must be carefully designed in order to guarantee convergence and optimality \cite{walrand-08,proutiere-10}. Similar problems arise for those adaptive CSMA algorithms where the  attempt rates are functions of the {\it queue lengths} instead of some slowly varying estimates of the arrival rates and service rates
\cite{srikant-10,raja-08}: the algorithm converges only for some specific choices of these functions.

In all these papers, optimality is defined either in terms of stability, assuming exogenous random packets arrivals at each link, or in terms of utility maximization, cf.~\cite{walrand-08,proutiere-10}. The flow-level dynamics are not considered, whereas they are key to understanding network performance  \cite{MR00}. In particular, it can be argued that the very notion of {\it congestion} should be defined at the flow level \cite{sigcomm}. In a recent paper, van de Ven, Borst and Shneer have shown that  the maximal weight scheduling policy, which is known to stabilize the network at the   packet level, may be unable to stabilize the network at  the flow level, which highlights the difference between the two notions of stability \cite{borst-09}. The main  contribution  of the present paper is to provide an algorithm that stabilizes the network {\it at flow level} whenever possible. With this objective in mind, it is very natural to think of  flow-aware CSMA. The fact that it suffices for each flow to run its own CSMA algorithm  is far from obvious, however. It is for instance well-known that maximizing the total throughput of the network at any time may make the network unstable at flow level  \cite{BM01}. It turns out that the fairness imposed by the proposed flow-aware CSMA is indeed sufficient to achieve stability.

Specifically, the flow-aware CSMA algorithm selects each feasible schedule in proportion to its  {\it weight}, where the weight of a schedule is the {\it product} of the number of flows on the corresponding links. For a large number of flows, the selected schedules are close to the corresponding {\it maximal weight} schedule (with product weights instead of additive weights), a policy that  turns out to be optimal. We note that a similar property is used by Ni, Bo and Srikant in \cite{srikant-10} for proving the stability of queue-length based CSMA at packet level. The constraints imposed by the packet level, like the
above mentioned problem of time-scale separation that restricts the set of eligible weight functions, make their algorithm very different from ours, however. Our model is purely asynchronous and stateless, the number of active flows at each link being determined by the packet headers in the corresponding buffer; moreover, the time-scale separation assumption is very natural in our case since  the attempt rates are adapted at the flow time-scale, which is typically much slower than the packet time-scale.

\section{Model}
\paragraph{Wireless network}
We consider the general model described in \cite{walrand-08}. There are  $K$ links in the network, where each link is an ordered transmitter-receiver pair. The network is associated with a conflict graph  $G=(V,E)$, where $V$ is the set of vertices (each representing a link) and $E$ is the set of edges (each representing a conflict). Two links $k,l$ can be simultaneously active if and only if they do not conflict, that is if $(k,l)\not \in E$. We refer to a {\it feasible schedule} as any set of  links  $S\subset V$ (possibly empty) that do not conflict with each other. We denote by $N$ the number of distinct feasible schedules and by $S_i$ the set of active links in schedule $i$, for all $i=1,\ldots,N$. By convention, schedule 1 corresponds to  the schedule where all nodes are idle, that is $S_1=\emptyset$.

\begin{figure}[h]
\begin{center}
\includegraphics[width=6cm]{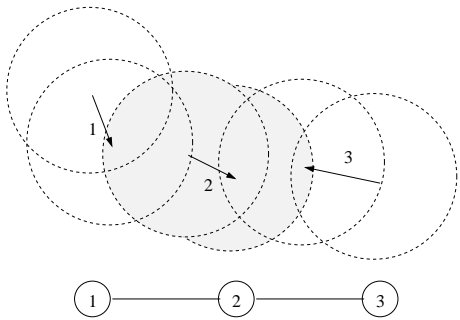}
\caption{\label{exclusion}A 3-link network and its conflict graph.}
\end{center}
\end{figure}

Consider  the network of $K=3$ links depicted by Fig.\ref{exclusion} for instance. Two links conflict if and only if the distance between the transmitter or receiver of one link and the transmitter or receiver of the other link is less than some fixed threshold. The conflict graph is linear and there are  $N=5$ feasible schedules, corresponding to the sets of active links $\emptyset, \{1\},\{2\},\{3\},\{1,3\}$.

\paragraph{Capacity region}
Let $\varphi_k$ be the physical rate of link $k$ when scheduled, in bit/s. The throughput of link $k$ when 
   each schedule $i$ is selected with probability $p_i$, with $\sumi p_i=1$, is given by:
\begin{equation}\label{eq:mu}
\forall k=1,\ldots,K, \quad\phi_k=\varphi_k \sum_{i:k\in S_i} p_i.
\end{equation}
Let  $\phi$ be the corresponding vector.
We refer to the {\it capacity region} as the set of vectors $\phi$ generated by all probability measures $p_1,\ldots,p_N$.

\paragraph{Flow-level dynamics}
Assume that flows arrive according to a Poisson process of intensity  $\lambda_k>0$ at link $k$ and have exponential flow sizes of mean $\sigma_k>0$, in bits.  
We denote by $\rho_k=\lambda_k\sigma_k$ the traffic intensity at link $k$  (in bit/s) and by $\rho$ the corresponding vector.
Let $x_k$ be the number of active flows at link $k$.  
We refer to the vector $x$ as the network state. 

We shall consider random access algorithms that select each schedule $i$ with some probability $p_i(x)$ that depends on the network state $x$, with $\sumi p_i(x)=1$. 
Under the time-scale separation assumption, the schedules change at a very high frequency compared to the flow-level time-scale, so that  the throughput   of link $k$ in state $x$ is given by:
\begin{equation}\label{eq:mux}
\phi_k(x)=\varphi_k \sum_{i:k\in S_i} p_i(x).
\end{equation}

The evolution of the network state then defines a Markov process $X(t)$  with transition rates $\lambda_k$ from state $x$ to state $x+e_k$ and $\mu_k(x)=\phi_k(x)/\sigma_k$ from state $x$ to state $x-e_k$ (provided $x_k>0$), where $e_k$ denotes the $K$-dimensional unit vector on component $k$.

\paragraph{Stability condition}
We are interested in the stability of the network in the sense of the positive recurrence of the Markov process $X(t)$. 
A necessary condition is that the vector traffic intensities $\rho$ lies in the capacity region. 
We look for distributed access schemes that stabilize the network whenever possible, that is for all vectors of traffic intensities $\rho$ in the interior of the capacity region. Such access schemes are referred to as {\it optimal}. For the sake of completeness, we first give an example showing the suboptimality of standard CSMA, that realizes some form of {\it maximal size} scheduling. 
We then prove the optimality of flow-aware CSMA.

\section{Standard CSMA}
\paragraph{Algorithm}
We first consider a standard CSMA algorithm where each link  waits for a period of random duration referred to as the {\it backoff time} before each transmission attempt. If the radio channel is sensed idle (in the sense that no conflicting link is active), a packet  is transmitted; otherwise, the link waits for a new backoff time  before the next attempt. Packets have random sizes of mean $\theta_k$ bits at link $k$ and are transmitted at the physical rate $\varphi_k$;
 the backoff times are random with mean $\tau_k$ at link $k$. We denote by $\alpha_k=\theta_k/(\varphi_k \tau_k)$ the ratio  of mean packet transmission time to mean backoff time at link $k$.

\paragraph{Equivalent scheduling}
We look for the steady-state probability $p_i(x)$ that the set of active links corresponds to schedule $i$ in state $x$.  We  assume that, in state $x$, each link $k$ such that $x_k>0$ takes all opportunities offered by the CSMA algorithm to transmit packets; any other link remains idle. If the packet sizes and the backoff times  had exponential distributions and there were no conflict, the evolution of the set of active links $S$ would form a reversible Markov process. A stationary measure of this Markov process is given by  1 if $S=\emptyset$ and:
 $$\prod_{k\in S}\alpha_k \I(x_k>0)$$
 otherwise.
 By reversibility, the actual stationary measure induced by the conflict graph is the truncation of this measure to the set of feasible schedules. Specifically, the weight $w_i(x)$ of feasible schedule $i$ in the stationary measure is given by:
$$
w_1(x)=1,\quad w_i(x)=\prod_{k\in S_i} \alpha_k \I(x_k>0)\quad \mbox{for all\ }i=2,\ldots,N.
$$
We deduce that schedule $i$ is selected in state $x$ with probability:
\begin{equation}\label{eq:pix}
p_i(x)={w_i(x)\over \sum_{j=1}^N w_j(x)}.
\end{equation}
By the insensitivity property of the underlying loss network, this is also the probability that schedule $i$ is selected in state $x$ for arbitrary phase-type distributions of packet sizes and backoff times with the same means; such distributions are known to form a dense subset within the set of all distributions with real, non-negative support \cite{bonald-07}.

\paragraph{Suboptimality}
We provide simple examples showing the suboptimality of the standard CSMA algorithm. We consider unit physical rates, that is $\varphi_k=1$ for all links $k$.
For a single link, the optimal stability condition is $\rho_1< 1$. In view of (\ref{eq:mux}) and (\ref{eq:pix}), the throughput is given by:
$$
\phi_1(x)={\alpha_1\over 1+\alpha_1}.
$$
We deduce the actual stability condition:
$$
\rho_1< {\alpha_1\over 1+\alpha_1}.
$$
This loss of efficiency is due to the backoff times, that must be chosen sufficiently small to limit the overhead of the CSMA algorithm.

Now consider  the example of Fig.~\ref{exclusion} with $K=3$ links. The optimal stability condition is given by:
$$
\rho_1+\rho_2< 1\et \rho_2+\rho_3< 1.
$$
Assume for simplicity all links have the same mean packet sizes and mean backoff times, so that $\alpha_1=\alpha_2=\alpha_3=\alpha$ for some $\alpha>0$. In view of (\ref{eq:mux}) and (\ref{eq:pix}), the throughput of the links in state $x$ are given by:
$$
\phi_1(x)=
\left\{
\begin{array}{ll}
{\alpha\over 1+\alpha}&\mbox{if\ }x_2=0, \\
{\alpha\over 1+2\alpha}&\mbox{if\ }x_2>0, x_3=0,\\
{\alpha+\alpha^2\over 1+3\alpha+\alpha^2}&\mbox{if\ }x_2>0, x_3>0,
\end{array}
\right.
$$
and
$$
\phi_2(x)=
\left\{
\begin{array}{ll}
{\alpha\over 1+\alpha}&\mbox{if\ }x_1=0, x_3=0,\\
{\alpha\over 1+2\alpha}&\mbox{if\ }x_1>0, x_3=0, \mbox{\ or \ } x_1=0, x_3>0,\\
{\alpha\over 1+3\alpha+\alpha^2}&\mbox{if\ }x_1>0, x_3>0.
\end{array}
\right.
$$
The throughput of link 3 follows by symmetry. As for a single link, the backoff times must be chosen sufficiently small to limit the overhead of the algorithm. In the limit $\alpha\to \infty$, we get:
\begin{equation}\label{eq:mu3}
(\phi_1(x),\phi_2(x),\phi_3(x))=\left\{
\begin{array}{ll}
(1,0,1)&\mbox{if\ }x_1>0, x_3>0,\\
(1/2,1/2,0)&\mbox{if\ }x_1>0, x_2>0, x_3=0, \\
(1,0,0)&\mbox{if\ }x_1>0, x_2=0, x_3=0, \\
(0,1,0)&\mbox{if\ }x_1=0, x_2>0, x_3=0,
\end{array}
\right.
\end{equation}
the other cases following by symmetry. Note that link 2 is not served when both
links 1 and 3 are active. This is due to the fact that link 2 is in conflict with
both links 1 and 3 and thus cannot access the channel for an infinitely small backoff time. This results in a suboptimal stability region:

\begin{prop}
The stability region is given by:
$$
\rho_1<\frac{1+\rho_3}{2},\ \rho_3<\frac{1+\rho_1}{2},\ \rho_2 < \pi_0+ \frac{\pi_{1,3}}{2},
$$
or
$$
\rho_1<\frac{1+\rho_3}{2},\ 
\frac{1+\rho_1}{2}\leq\rho_3 < \frac{1+\rho_1}{2}+ {1-\rho_1\over 2}\pi_{2,1},\ \rho_2< \frac{1-\rho_1}{2},
$$
or
$$
\rho_3<\frac{1+\rho_1}{2},\ 
\frac{1+\rho_3}{2}\leq\rho_1 < \frac{1+\rho_3}{2}+ {1-\rho_3\over 2}\pi_{2,3},\
\rho_2< \frac{1-\rho_3}{2},
$$
where $\pi_0$, $\pi_{1,3}$, $\pi_{2,1}$ and $\pi_{2,3}$ are the respective
probabilities that:
\begin{itemize}
\item both links
$1$ and $3$ are idle when link $2$ is always active;
\item one of the links
$1$ or $3$ is idle when link $2$ is always active;
\item  link $2$ is idle given that link $1$ is idle, when link $3$ is always active;
\item  link $2$ is idle given that link $3$ is idle, when link $1$ is always active.
\end{itemize}
More precisely, the Markov process $X(t)$ is positive recurrent if the vector 
 of traffic intensities $\rho$ lies in this region and transient if it lies outside its closure. 
\end{prop}

\begin{figure}[h]
\begin{center}
\includegraphics[width=8cm]{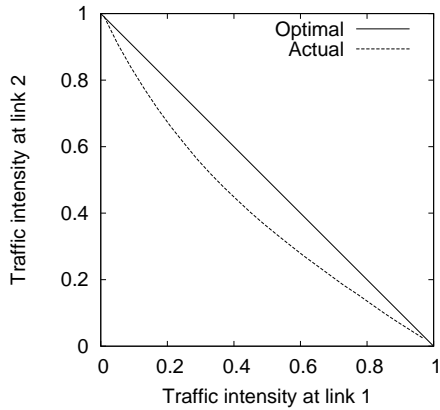}
\caption{\label{stab}Stability condition for the network of Fig. 1 under standard CSMA ($\rho_1=\rho_3$).}
\end{center}
\end{figure}

The proof is given in the Appendix. Note that, when one of the links is always active, the two other links form 
a coupled system of two queues as considered by Fayolle and Iasnogorodski
\cite{fayolle-79}. In particular, the stability region can be calculated exactly. 
In the symmetric case $\rho_1=\rho_3$, the stability 
condition reduces to
$
\rho_1 < 1,\  \rho_2 <\pi_0 + \pi_{1,3}/{2}.
$
Fig. \ref{stab} shows that the corresponding stability region for equal mean flow sizes.

\section{Flow-aware CSMA}
\paragraph{Algorithm}
We now consider the flow-aware CSMA algorithm where each {\it flow} (instead of each link) waits for a random backoff time before each transmission attempt. If the radio channel is sensed idle (in the sense that no conflicting link is active, nor any other flow on the same link), a packet of this flow is transmitted; otherwise, the flow remains idle for a new random backoff time before the next attempt. The backoff times have random durations of mean $\tau_k$  for each active flow at link $k$. We still denote by $\alpha_k=\theta_k/(\varphi_k \tau_k)$ the ratio  of mean packet transmission time to mean backoff time at link $k$.

\paragraph{Equivalent scheduling}
Again, we look for the steady-state probability $p_i(x)$ that the set of active links corresponds to schedule $i$ in state $x$.  We assume that all active flows take each opportunity offered by the CSMA algorithm to transmit packets. 
 If the packet sizes and the backoff times  had exponential distributions and there were no conflict,
the evolution of the set of active links $S$   would again form a reversible Markov process. Since there are $x_k$ flows attempting to access the channel at link $k$, a stationary measure of this Markov process is given by  1 if $S=\emptyset$ and:
$$\prod_{k\in S}\alpha_k x_k$$ 
otherwise.
 By reversibility, the actual stationary measure induced by the conflict graph is the truncation of this measure to the set of feasible schedules. The weight $w_i(x)$ of feasible schedule $i$ in the stationary measure is given by:
$$
w_1(x)=1,\quad w_i(x)=\prod_{k\in S_i} \alpha_k  x_k\quad \mbox{for all\ }i=2,\ldots,N.
$$
Schedule $i$ is then selected with probability $p_i(x)$ given by (\ref{eq:pix}) in state $x$.
By the insensitivity property of the underlying loss network,  this probability remains the same for arbitrary phase-type distributions of packet sizes and backoff times  with the same means, cf. \cite{bonald-07}.

\paragraph{Optimality}
We now give the main result of the paper, that demonstrates the optimality of the above flow-aware CSMA algorithm. 

\begin{theorem}
The network is stable for all vectors of traffic intensities $\rho$ in the interior of the capacity region. 
\end{theorem}
\bp
We apply Foster's criterion. Specifically, we look for some Lyapunov function $F(x)$ such that the corresponding drift, given  by:
$$
\Delta F(x)=\sum_{k=1}^K \lambda_k (F(x+e_k)-F(x))+ \sum_{k:x_k>0} \mu_k(x) (F(x-e_k)-F(x)),
$$
satisfies:
$$
\Delta F(x)\le -\delta
$$
for some $\delta>0$, in all states $x$ but some finite number. 

If the vector of traffic intensities $\rho$ lies in the interior of the capacity region, there exists some $\epsilon>0$ and some probability measure $q_1,\ldots,q_N$ on the set of feasible schedules such that $q_i>0$ for all $i=1,\ldots,N$ and:
\begin{equation}\label{eq:stab}
\forall k=1,\ldots,K,\quad\rho_k=(1-2\epsilon)\varphi_k\sum_{i:k\in S_i}q_i.
\end{equation}
Define:
$$
F(x)=\sum_{k:x_k>0} {\sigma_k\over \varphi_k}x_k(\log(\alpha_kx_k)-1).
$$
We get:
$$
\Delta F(x)=G(x) +\sum_{k:x_k>0}{\rho_k\over \varphi_k}(x_k+1)(\log(1+{1\over x_k})-1)
$$
\begin{equation}\label{dfx}
+\sum_{k:x_k>0}{\phi_k(x)\over \varphi_k}(x_k-1)(\log(1-{1\over x_k})+1),
\end{equation}
with:
$$
G(x)=\sum_{k:x_k>0}{\rho_k-\phi_k(x)\over \varphi_k}\log(\alpha_kx_k).
$$
Noting that, for any probability measure $p_1,\ldots,p_N$ on the set of feasible schedules:
$$
\sum_{k:x_k>0} \sum_{i:k\in S_i}p_i \log (\alpha_kx_k)=\sumi p_i \log(w_i(x)),
$$
we get using (\ref{eq:stab}):
$$
G(x)=-{\epsilon} \sumi q_i \log(w_i(x))+\sumi (q_i(1-{\epsilon})-p_i(x))\log(w_i(x)).
$$
We then need the following lemma.

\begin{lemma}
Let:
$$
w(x)=\max_{i=1,\ldots,N}w_i(x).
$$
Then, for all states $x$ but some finite number,
$$
\sumi p_i(x)\log(w_i(x))\ge (1-\epsilon) \log(w(x)).
$$
\end{lemma}
\bp
The proof is similar to that of \cite[Proposition 2]{srikant-10}.
Let:
 $$I(x)=\left\{i=1,\ldots,N: \log(w_i(x))\ge  (1-{\epsilon\over 2}) \log(w(x))\right\}.$$
We have:
$$
\sumi p_i(x)\log(w_i(x))\ge (1-{\epsilon\over 2}) \log(w(x))\sum_{i\in I(x)} p_i(x).
$$
Moreover,
\begin{eqnarray*}
\sum_{i\not \in I(x)} p_i(x)&=&  {\sum_{i\not \in I(x)} w_i(x)\over \sumi w_i(x)}, \\
&\le & {(N-|I(x)| ) w(x)^{1-{\epsilon\over 2}}\over  w(x)}, \\
&=& {N-|I(x)| \over  w(x)^{\epsilon\over 2}}.
\end{eqnarray*}
Since $w(x)$ tends to $+\infty$ when $|x|=\sumk x_k$ tends to  $+\infty$, this quantity 
is less than $\epsilon/2$ for all states $x$ but some finite number.
We deduce that in all states $x$ but some finite number:
$$
\sumi p_i(x)\log(w_i(x))\ge (1-{\epsilon\over 2})^2 \log(w(x))\ge (1-{\epsilon}) \log(w(x)).
$$
\ep
\\

In view of Lemma 1, we have for all states $x$ but some finite number:
$$
G(x)\le -{\epsilon} \sumi q_i \log(w_i(x))+(1-{\epsilon})\sumi (q_i \log(w_i(x))-\log(w(x))).
$$
Since $w_i(x)\le w(x)$ for all states $x$, we deduce that  for all states $x$ but some finite number:
$$
G(x)\le -{\epsilon} \sumi q_i \log(w_i(x)).
$$
Since $q_i>0$ for all $i=1,\ldots,N$, this expression tends to $-\infty$ when  $|x|=\sumk x_k$ tends to  $+\infty$.
The other terms of $\Delta F(x)$ in (\ref{dfx}) being bounded,  we deduce that there exists $\delta>0$ such that $\Delta F(x)\le -\delta$ for all states $x$ but some finite number.
\ep

\section{Throughput performance}
\label{sec:perf}

This section is devoted to the throughput performance of flow-aware CSMA, under the stability condition.
We are interested in the {\it mean throughput}, defined as the ratio of the mean flow size  to the mean flow duration. By Little's law, the mean throughput at link $k$ is given by:
\begin{equation}\label{gammak}
\gamma_k={\rho_k\over \mbox{E}[x_k]}.
\end{equation}
We consider unit physical rates, that is $\varphi_k=1$ for all links $k$.

\begin{figure}[h]
\begin{center}
\includegraphics[width=8cm]{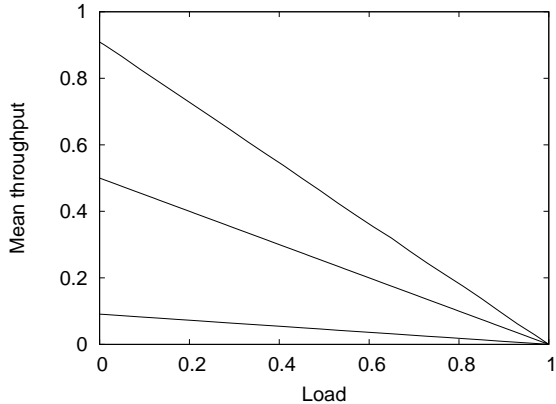}
\caption{\label{single}Impact of the mean backoff time on the mean throughput for a single link (ratio of mean packet transmission time to mean backoff time $\alpha_1=0.1,1,10$, from bottom to top).}
\end{center}
\end{figure}

\paragraph{Single link} We first analyse the impact of the mean backoff time on the mean throughput in the case of a single link. In the presence of $x_1$ flows,  the total throughput is given by:
$$
\phi_1(x_1)={\alpha_1 x_1\over 1+\alpha_1x_1}.
$$
The number of flows then behaves as the number of customers in a processor-sharing queue with state-dependent service rate. 
The corresponding stationary distribution is given by:
$$
\pi(x_1)=\pi(0)\prod_{n=1}^{x_1}{\rho_1\over \phi_1(n)},
$$
under the stability condition $\rho_1<1$. The mean throughput then follows from (\ref{gammak}).
For $\alpha_1\to \infty$, the throughput is constant and equal to 1  and the mean throughput is given by $\gamma_1=1-\rho_1$; for $\alpha_1=1$, the  system corresponds to a processor-sharing queue with an additional permanent customer representing the  backoff times and we have $\gamma_1=(1-\rho_1)/2$; in general, we have $\gamma_1\to \alpha_1/(1+\alpha_1)$ when $\rho_1\to 0$. These results are illustrated by Fig.~\ref{single}.

\begin{figure}[h]
\begin{center}
\includegraphics[width=8cm]{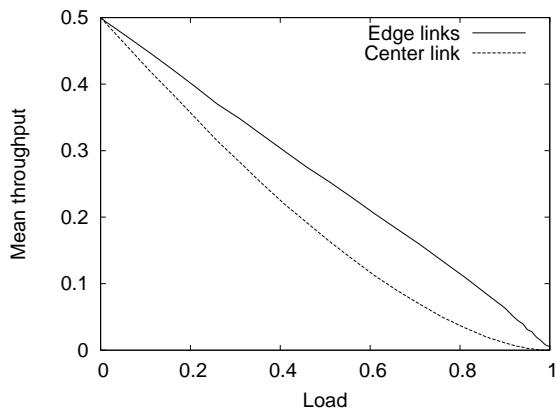}
\caption{\label{line3}Mean throughput in the 3-link line.}
\end{center}
\end{figure}

\begin{figure}[h]
\begin{center}
\includegraphics[width=7cm]{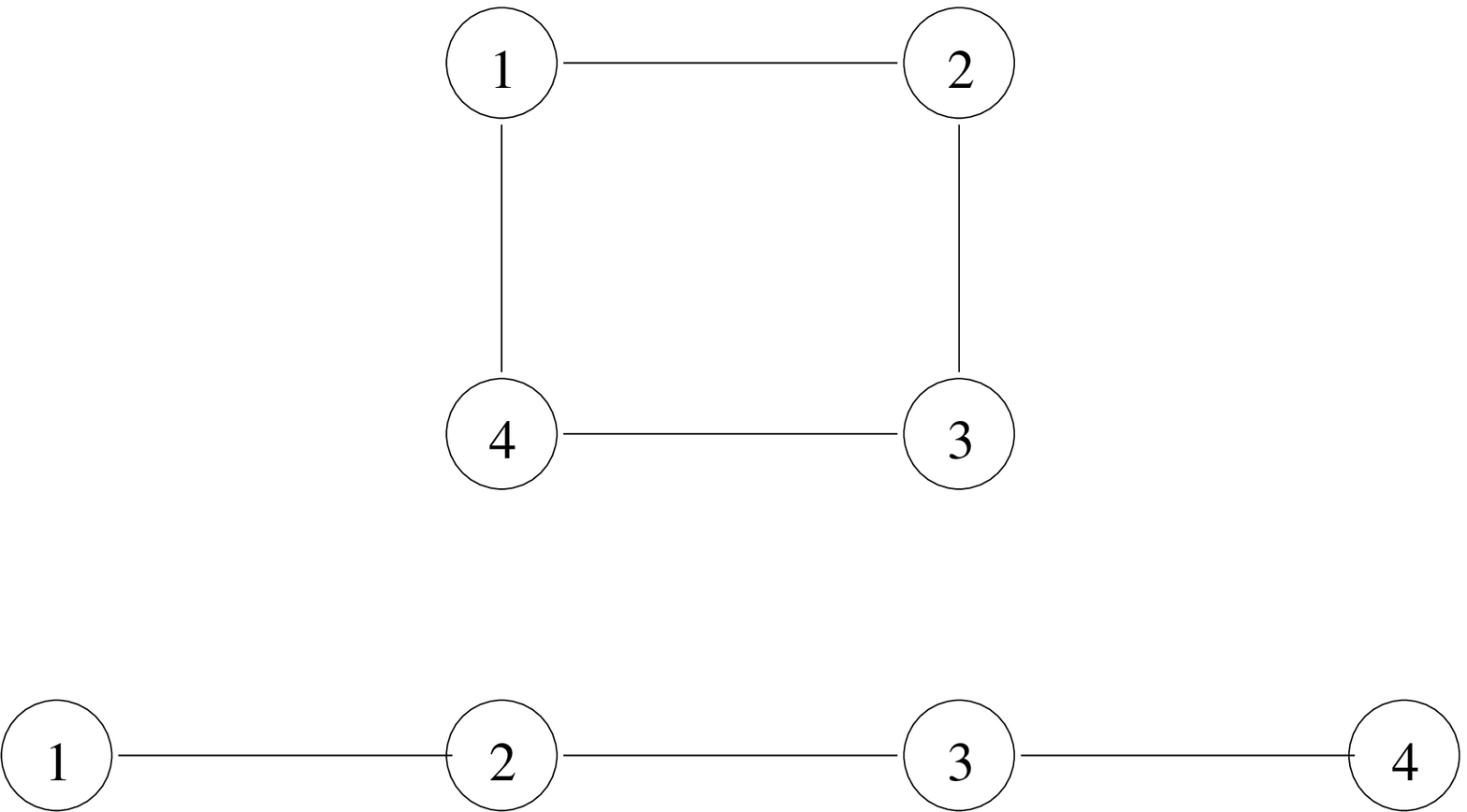}
\hspace{1cm}
\includegraphics[width=3.5cm]{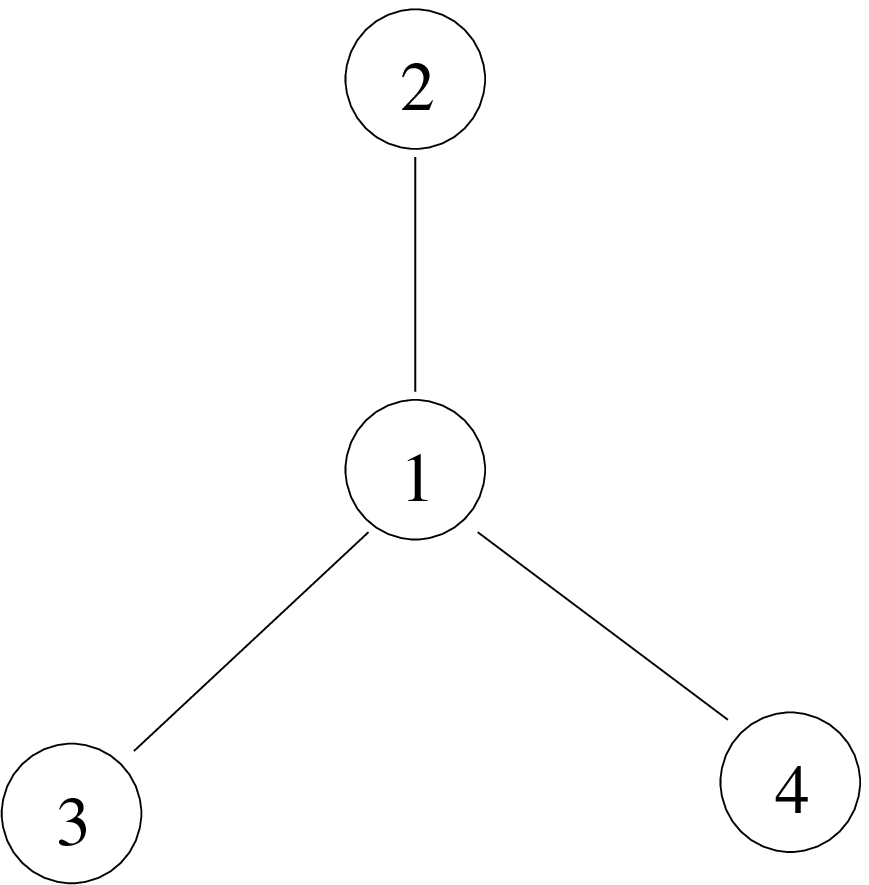}
\caption{\label{square-net}Conflict graphs of the square, the 4-link line and the 4-link star.}
\end{center}
\end{figure}

\begin{figure}[h]
\begin{center}
\subfigure[Square and  line]{\includegraphics[width=8cm]{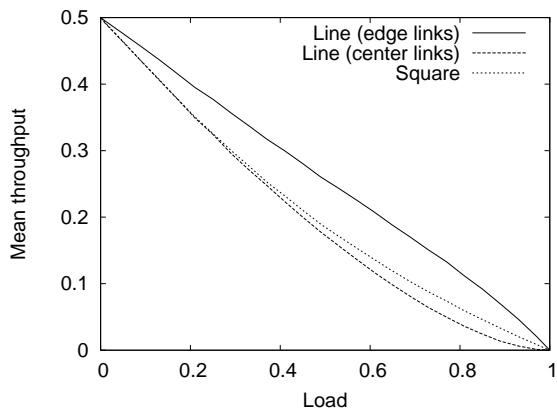}}\\
\subfigure[Star]{\includegraphics[width=8cm]{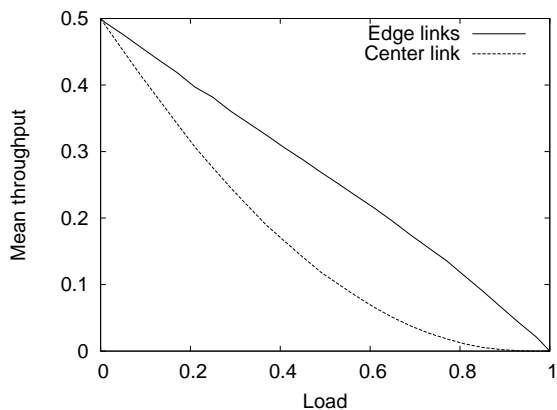}}
\caption{\label{square}Mean throughput in 4-link networks.}
\end{center}
\end{figure}

\paragraph{Networks} In the following, we consider network scenarios and assume that the mean backoff time is  the same for all flows and equal to the mean packet transmission time, so that $\alpha_k=1$ for all links $k$.  Flows have unit mean flow sizes. The traffic intensity is the same on all links, equal to $\rho_1$.  We refer to the network load as the ratio of the per-link traffic intensity $\rho_1$ to its maximum value, given by the stability condition. Fig.~\ref{line3} and \ref{square} give the results obtained  for the 3-link line of Fig.~\ref{exclusion} and for the three 4-link networks of Fig.~\ref{square-net}, for the same mean flow sizes . The results are obtained by the simulation of $10^7$ jumps of the underlying Markov process, after a warm-up period of $10^5$ jumps. We observe that the throughput decreases from its maximum value $1/2$ to 0 when the load grows from 0 to 1; it is lower on links that are in conflict with many other links, just like in wired networks, the mean throughput is lower  on long routes, where flows go through many links \cite{BM01}.

\section{Conclusion}
\label{sec:conclusion}

The standard CSMA algorithm is inherently unfair and inefficient. We have shown that the proposed flow-aware CSMA algorithm, where each {\it flow} (instead of each link) runs its own CSMA algorithm, is not only fair but efficient, in the sense that the network is stable whenever possible. To our knowledge, this is the first distributed algorithm that is provably optimal in terms of flow-level stability. 

The considered packet-level model relies on a number of simplifying assumptions
that we plan to relax in future work. These include the absence of collisions and  hidden nodes. The interaction with the usual back-off mechanism of IEEE 802.11 should also be studied. One may also envisage different implementations of the  proposed flow-aware CSMA algorithm where the attempt rate of each link is equal to some increasing function of the number of flows and the transmission opportunities are shared in a fair way between active flows, using a deficit round-robin scheduler for instance. 

 From a more theoretical perspective, it would be worth relaxing the assumption of exponential flow sizes
 and deriving bounds or approximations on the throughput performance of the algorithm.

\appendix

\section*{Appendix}

\section*{Proof of Proposition 1}
This example is similar to the one studied in \cite[p274]{robert-03}.
We consider the fluid limits of the Markov process $X(t)$. Specifically,
we define $X^{(n)}(t)$ as the Markov process $X(t)$ whose initial state
is $X^{(n)}(0)=(\lfloor\beta_1 n\rfloor,\lfloor\beta_2 n\rfloor,\lfloor\beta_3 n\rfloor)$
for some non-negative real numbers $\beta_1,\beta_2, \beta_3$ such that
$\beta_1+\beta_2+\beta_3=1$.
We then define:
$$
\bar{X}^{(n)}(t)= \frac{1}{n} X^{(n)}(nt).
$$
The fluid limits of the Markov process $X(t)$, if they exist, are the limiting
points of this set of processes when $n\to +\infty$.
It is easy to check that the Markov process $X(t)$ belongs to the class $(\mathcal{C})$\
defined in \cite[p241]{robert-03} and that the associated Proposition 9.3 applies.
In particular, the set $\{\bar{X}^{(n)}(t), n\in \N\}$ is tight and the fluid
limits are continuous. The Markov process  $X(t)$ is then positive recurrent if
there exists some finite time  after which all fluid limits are null,
cf.~\cite[Theorem 9.7, p259]{robert-03}; it is
transient if there exists some initial state $\beta_1,\beta_2,\beta_3$ such 
that, after  some finite time, some components of the fluid limits grow
at least linearly to infinity \cite{meyn-95}.

We first calculate   the fluid limit until the first time where one component
reaches 0, if any, for all possible initial states. The three components  of the process
$X^{(n)}(t)$ behave as three coupled $M/M/1$ queues, with arrival rates $\lambda_1,\lambda_2,\lambda_3$ and state-dependent service rates. We denote by 
 $\mu_k=1/\sigma_k$ the maximum service rate of queue $k$, so that $\rho_k=\lambda_k/\mu_k$. The Markov process is positive recurrent if all queues empty in finite time in the limit and transient if, starting from some initial state, at least one queue grows linearly to infinity after some finite time.
 
 We start with  the case
$\beta_1>0$, $\beta_2>0$, $\beta_3>0$. The three queues are then mutually independent, with respective service rates $\mu_1,0,\mu_3$.
The scaling property of the $M/M/1$ queue shows that the process $\bar{X}^{(n)}(t)$
weakly converges to the function:
$$(\beta_1+(\lambda_1-\mu_1)t,\beta_2+\lambda_2t,\beta_3+(\lambda_3-\mu_3)t),$$
until one of the components reaches 0, if any.

We now consider the case  $\beta_1=0$, $\beta_2>0$, $\beta_3>0$. In view of
(\ref{eq:mu3}), queue 1 has service rate $\mu_1$ and is empty with probability $1-\rho_1$.
Queues 2 and 3 have service rates
$0$, $\mu_3$ with probability $\rho_1$ and $\mu_2/2$, $\mu_3/2$ with probability $1-\rho_1$.
Proposition 9.14 of \cite{robert-03} applies and the process $\bar{X}^{(n)}(t)$
weakly converges to the function:
$$
(0,\beta_2+(\lambda_2 - \mu_2\frac{1-\rho_1}{2})t,\beta_3+(\lambda_3 - \mu_3\frac{1+\rho_1}{2})t),
$$
until one of the components reaches 0, if any.

Next, we consider the case  $\beta_1=\beta_2=0$,  $\beta_3>0$.
In view of (\ref{eq:mu3}), queue $1$ has service rate $\mu_1$. Queue $2$ has service rate $\mu_2/2$ if queue 1 is empty and $0$ otherwise.
This queue is stable if $\rho_2< (1-\rho_1)/2$, which we assume.
Queue 2 then remains empty in the limit, and
the service rate of queue 3 is
 equal to $\mu_3$ with probability
$\rho_1+(1-\rho_1)\pi_{2,1}$ and to $\mu_3/2$ otherwise.
 We deduce that the process $\bar{X}^{(n)}(t)$ weakly
converges to the function:
$$
(0,0,\beta_3+(\lambda_3-\mu_3({\rho_1+1\over 2}-{1-\rho_1\over 2}\pi_{2,1}))t),
$$
whenever component 3 is positive.

Finally, we consider the case  $\beta_1=\beta_3=0$, $\beta_2>0$.
In view of (\ref{eq:mu3}), the service rates of queues $1$ and $3$ are equal to  $\mu_1$ and $\mu_3$ when both are non-empty and to
$\mu_1/2$ and $\mu_3/2$ otherwise. This system is stable if $\rho_1<(1+\rho_3)/2$ and
$\rho_3< (1+\rho_1)/2$, which we assume. Queues 1 and 3 then remain
empty in the limit. The service rate of queue 3 is equal to $\mu_2$ with probability $\pi_0$ and to $\mu_2/2$ with probability $\pi_{1,3}$.
The  process $\bar{X}^{(n)}(t)$
weakly converges to the function:
$$
(0,\beta_2+(\lambda_2-\mu_2(\pi_0-\frac{\pi_{1,3}}{2}))t,0),
$$
whenever component 2 is positive.

To conclude the proof,
we consider the evolution of the  fluid limit in the following five cases (the others follow by symmetry):
\begin{enumerate}
\item
Assume $\rho_1<(1+\rho_3)/2$ and $\rho_3<(1+\rho_1)/2$. Note that 
this implies $\rho_1<1$ and $\rho_3<1$.
Queue 1 and 3 empty in finite time, independently 
of queue 2.  Queue $2$ then empties  in finite time if 
$\rho_2 < \pi_0+\pi_{1,3}/2$; it 
grows  linearly to infinity if $\rho_2 > \pi_0+\pi_{1,3}/2$.

\item  Assume $\rho_1<(1+\rho_3)/2$ and $\rho_3>(1+\rho_1)/2$. 
If $\rho_1 \geq 1$ then $\rho_3>1$ and queue 3 grows linearly to infinity.
We now assume $\rho_1<1$.  If $\rho_2> (1-\rho_1)/2$ then  queue 2 grows  linearly 
to infinity. 
If $\rho_2 = (1-\rho_1)/2$ then starting from a state where $\beta_1=0$, 
$\beta_2>0$ and $\beta_3>0$, queue 1 stays empty, queue 2 is constant and queue 3 grows
linearly to infinity. We assume that  $\rho_1< 1$ and $\rho_2< (1-\rho_1)/2$.
Starting from the initial state $\beta_1=\beta_2=0$, $\beta_3>0$, queue 3 grows
 linearly to infinity if 
$\rho_3 >({1+\rho_1})/{2}+ \pi_{2,1}({1-\rho_1})/ 2$. We assume that $\rho_3 <({1+\rho_1})/{2}+ \pi_{2,1}({1-\rho_1})/ 2$. 
Starting from the initial state $\beta_1=\beta_2=0$, $\beta_3>0$, queue 
3 then empties in finite time. It remains to prove that, starting from any 
initial state, queues 1 and 2 empty in finite time. We first note that, 
since $\rho_1<1$ and $\rho_3<1$,   queue 1 or queue 3 empties in finite time. 
Moreover, if both queues 1 and 3 are empty but not queue  2,
 then queue 3 grows linearly. Thus we can assume that queue 1 
empties before queue 3. We know that queue 2 empties in finite time 
in this case. 

\item Assume  $\rho_1<(1+\rho_3)/2$ and $\rho_3=(1+\rho_1)/2$. 
Note that $\rho_1<1$ and $\rho_3<1$ in this case. Moreover, we have 
$\pi_0=0$ and $\pi_{1,3}=1-\rho_1$, so that the inequality $\rho_2< 
\pi_0+\pi_{1,3}/2$ is equivalent to $\rho_2<(1-\rho_1)/2$. 
If the latter is satisfied, then if queue 1 is non-empty then queue 2
empties in finite time independently of queue $3$. We just have
to consider the case where $\beta_1=\beta_2=0$ and $\beta_3>0$. Because 
$\rho_3=(1+\rho_1)/2<({1+\rho_1})/{2}+ \pi_{2,1}({1-\rho_1})/ 2$,
queue 3 empties in finite time.
If $\rho_2>(1-\rho_1)/2$, we choose an initial state such that queue $1$ empties 
before $3$. When queue 1 is empty, queue 3 is constant and queue
2 grows linearly to infinity.

\item Assume  $\rho_1\ge (1+\rho_3)/2$ and $\rho_3> (1+\rho_1)/2$. 
Then $\rho_1>1$ and $\rho_3>1$ so that queues 1 and  3 grow 
linearly to infinity.

\item  Assume  $\rho_1= (1+\rho_3)/2$ and $\rho_3= (1+\rho_1)/2$. 
Then $\rho_1=\rho_3=1$ and $\pi_0=\pi_{1,3}=0$. If $\rho_2=0$, the 
vector $\rho$ lies on the boundary of the stability region. If $\rho_2>0$,
queue 2  grows  linearly to infinity.
\end{enumerate}

\section*{References}
\bibliographystyle{elsarticle-harv}
\bibliography{biblio.bib}

\end{document}